\documentclass[aps,preprint,showpacs,superscriptaddress,groupedaddress]{revtex4-1}  
\usepackage{graphicx}  
\usepackage{dcolumn}   
\usepackage{bm}        
\usepackage{amssymb}   
\usepackage{amsmath}   
\usepackage{listings}
\usepackage{array}
\usepackage{graphicx}
\usepackage{lineno}
\usepackage{dcolumn}
\usepackage{color}
\usepackage{overpic}
\usepackage{multirow}
\usepackage{hyperref}

\newcommand{\hatds}{\hat{\sigma}_s}
\newcommand{\hatdmu}{\hat{\sigma}_{\mu}}
\newcommand{\hatda}{\hat{\sigma}_{\alpha}}
\newcommand{\hatdg}{\hat{\sigma}_{\gamma}}
\newcommand{\mL}{\mathcal{L}}
\newcommand{\hatmu}{\hat{\mu}}
\newcommand{\hats}{\hat{s}}
\newcommand{\hata}{\hat{\alpha}}
\newcommand{\hatg}{\hat{\gamma}}
\newcommand{\hatth}{\hat{\theta}}
\newcommand{\bV}{\bold{V}}
\newcommand{\bA}{\bold{A}}
\newcommand{\bI}{\mathbf{1}}
\newcommand{\bx}{\bold{x}}
\newcommand{\bX}{\bold{X}}
\newcommand{\ba}{\bold{a}}
\newcommand{\bb}{\bold{b}}
\newcommand{\bB}{\bold{B}}
\newcommand{\bC}{\bold{C}}
\newcommand{\bD}{\bold{D}}

\begin{document}

\widetext


\title{Study of constraint and impact of a nuisance parameter in maximum likelihood method}
\author{Li-Gang Xia \\ Department of Physics, Warwick University, CV4 7AL, UK}

\begin{abstract}
   Maximum likelihood method is widely used for parameter estimation in high energy physics. To consider various systematic uncertainties, tens of or even hundreds of nuisance parameters (NP) are introduced in a likelihood fit. The constraint of a nuisance parameter and its impact on the parameter of interest (POI) will be the main concerns for a precise measurement. A fit involving many parameters is usually slow and it is even more time-consuming to investigate why a parameter is over-constrained or has a large impact. In this paper, we are trying to understand the reasons behind and provide simple formulae to estimate the constraint and impact directly. 
\end{abstract}

\maketitle

\section{Introduction}
Maximum likelihood method is widely used in high energy physics, such as the observation of the Higgs boson~\cite{higgs_observation_atlas, higgs_observation_cms} at the Large Hadron Collider (LHC). Typically in a measurement by ATLAS or CMS collaboration, we have to estimate tens of or even hundreds of systematic sources. For example, 214 nuisance parameters are involved in the measurement of higgs properties in the diphoton decay channel~\cite{higgs_diphoton}. They may affect the normalization or/and shape of the observable distribution differently. In practice, we introduce a nuisance parameter (NP) for each systematic uncertainty in the likelihood function. Due to many fitting parameters, it may take hours for one fit and even more time to understand the potential features presented in the fitting results. For example, the post-fit uncertainty for a systematic source may be much smaller than its initial estimation in a measurement. In other words, the corresponding nuisance parameter turns out to be over-constrained in the fit. Then the measurement may be aggressive as the fit does not consider the full uncertainty. On the other side, we care about which systematic sources have large contribution to the uncertainty of the parameter of interest (POI). This is important if the data statistics is not the main limiting factor. In this paper, we are trying to understand why a parameter could be over-constrained or have a large impact on POI uncertainty. Meanwhile, we also present simple formulae to estimate the constraint and impact directly. It should be noted that advanced numerical methods have been developed to estimate them precisely. Our formulae will never be a substitute, but help us understand the physics reasons behind the fitting features. 

In Section~\ref{sec:simple_model}, we start with a simple model based on number-counting experiments and introduce the definition of constraint and impact for a nuissance parameter. It is extended to a more realistic model in Section~\ref{sec:realistic_model}. A toy experiment is performed for illustration in Section~\ref{sec:example}. The conclusion will be summarized in Section~\ref{sec:summary}.

\section{\label{sec:simple_model}A simple model}
Considering an experiment of counting number of events, let $n$ be the observed number of events, $b$ be the number of background events from Monte Carlo (MC) simulation and $s$ be the number of signal events to be determined. For the background prediction, let $\delta$ be the MC statistical uncertainty, and we introduce one systematic uncertainty $\Delta$. The likelihood function in this model is
\begin{equation}\label{eq:L_simple_model}
   \mL(s,\alpha,\gamma) = P(n|s+\gamma b + \alpha\Delta)\times P(m | \gamma m)\times G(\alpha|0,1) \:,
\end{equation}
where $s$ is the parameter of interest, which is the measurement target and can be used to discriminate the right theory model; $\alpha$ and $\gamma$ are two nuisance parameters as explained below; $m\equiv \frac{b^2}{\delta^2}$ is a constant; $P(n|\lambda) = \frac{\lambda^{n}}{n!}e^{-\lambda}$ is the Poisson distribution function with the expectation value $\lambda$, $G(x|\mu,\sigma) = \frac{1}{\sqrt{2\pi}\sigma}e^{-\frac{(x-\mu)^2}{2\sigma^2}}$ is the Gaussian distribution function with mean $\mu$ and standard deviation $\sigma$. The right-hand side of the equation is a
product of three factors. The first one is the Poisson probability of observing $n$ events with the expectation $s+\gamma b+\alpha\Delta$. We start with the expectation $s+b$. The second factor is to account for the background MC statistical uncertainty. It is considered by introducing a nuissance parameter $\gamma$ and an auxillary experiment with the observed number of events $m$ so that the original relative uncertainty $\frac{\delta}{b}$ is preserved. Thus the expectation $s+b$ is replaced by $s+\gamma b$. The third factor is
to include the estimation of the systematic uncertainty $\Delta$. This is done by introducing a nuisance parameter $\alpha$ which abides by the Gaussian distribution function with mean value 0 and variance 1 and replacing $s+\gamma b$ by $s+\gamma b + \alpha\Delta$.  The Gaussian constraint is generally used across the measurements~\cite{atlas1,atlas2,cms1,cms2} by ATLAS and CMS collaborations. Ignoring the irrelevant constant terms, the log likelihood function is then
\begin{equation}
   \ln \mL = n \ln(s+\gamma b + \alpha \Delta) - (s+ \gamma b +\alpha\Delta) -  \frac{\alpha^2}{2} + m \ln\gamma - \gamma m \:.
\end{equation}
Maximizing the likelihood function leads to the following estimation. 
\begin{equation}
   \hats = n-b\:, \quad \hata = 0\:,\quad \hatg = 1\: .
\end{equation}
Here a hat $\hat{}$ is added to represent the best-fit values.

Letting $\bV$ denote the covariance matrix of the fitting parameters, the inverse of its estimation is related with the second-order derivatives of the log likelihood function evaluated at the best-fit values, as shown in the following equation~\cite{book:glen_cowan}.
\begin{equation}
   (\bV^{-1})_{ij} = - \frac{\partial^2 \ln \mL (\hatth_i,\hatth_j)}{\partial \theta_i\partial\theta_j}
\end{equation}
where $\theta_i$s denote the parameters $(\theta_1,\theta_2,\theta_3)=(s,\alpha,\gamma)$. For this model, the inverse of $\bV$ is 
\begin{equation}
   \bV^{-1} = \frac{1}{n}
   \begin{pmatrix}
	1 & \Delta & b\\
	\Delta &  \Delta^2 + n & \Delta b \\
	b & \Delta b& b^2+n\frac{b^2}{\delta^2}\\
   \end{pmatrix} \: , 
\end{equation}
with the determinant $\det|\bV^{-1}| = \frac{b^2}{n\delta^2}$. The diagonal elements of $\bV$ give the uncertainty of the fitting parameters.
\begin{equation}~\label{eq:simple_model}
   \hatds =\sqrt{\bV_{11}}= \sqrt{n + \Delta^2 + \delta^2 }\:, \quad \hatda=\sqrt{\bV_{22}}= 1 \:, \quad \hatdg=\sqrt{\bV_{33}} = \frac{\delta}{b} \: .
\end{equation}

Based on the results above, let us introduce the definition of constraint and impact for a nuisance parameter studied in this paper. In this example, the nuisance parameter $\alpha$ corresponding to the systematic uncertainty $\pm \Delta$ is not over-constrained as $\hatda=1$ where 1 is our initial estimation. We define the constraint of a nuisance parameter as the ratio of the fitted variation to the input variation (unity by construction), namely, $\frac{\hatda}{1}=\hatda$. If $\hatda$ is much smaller than 1, we say $\alpha$ is over-constrained and we may worry because the uncertainty $\pm\Delta$ is not fully considered. On the other side, if $\hatda$ is higher than 1, it means larger uncertainty than initial estimation is considered. The latter case is usually not of our concern because the measurement is conservative.  The POI uncertainty can be expressed as a quadrature sum, namely, $\hatds^2 = \sqrt{n}^2 + \Delta^2 + \delta^2$. It has three parts, which represent the contribution from the data statistics, the systematic uncertainty ($\alpha$) and the MC statistical uncertainty ($\gamma$), respectively. Taking $\hatds$ as a function of $\Delta$ and $\delta$, the impact on the POI ( $s$ in this model ) of the nuisance parameter $\alpha$, denoted by $\hatds^{\alpha}$, can be defined as $\hatds^{\alpha}\equiv \sqrt{\hatds^2(\Delta,\delta)-\hatds^2(0,\delta)}=\Delta$. Similarly, the impact of $\gamma$ can be defined as $\hatds^{\gamma}\equiv\sqrt{\hatds^2(\Delta,\delta) - \hatds^2(\Delta,0)}=\delta$.

\section{\label{sec:realistic_model}A realistic model}
Extending the model above to a case that a measurement is performed in distributions of an observable, we resort to the binned likelihood estimation and the likelihood function becomes
\begin{equation}\label{eq:L_realistic_model}
   \mL(\mu,\alpha,\gamma) = \Pi_{i=1}^{N}P(n_i|\mu s_i+\gamma_i b_i + \alpha \Delta_i) P(m_i | \gamma_i m_i) \times G(\alpha|0,1) \:,
\end{equation}
where $N$ is the number of bins; $n_i$ is the observed number of events in the $i$-th bin while $s_i$ and $b_i$ are signal and background prediction; $\mu$ is the signal strength with respective to the prediction and is the parameter of interest; $\Delta_i$ is the systematical variation at the $i$-th bin with the corresponding nuisance parameter $\alpha$; $m_i\equiv \frac{b_i^2}{\delta_i^2}$ with $\delta_i$ being the MC statistical uncertainty at $i$-th bin and $\gamma_i$ being the
corresponding nuisance parameter. 

Differently from previous model, here the signal strength $\mu$ is fitted to give the the signal magnitude while the signal shape is determined by the theory model or the well-known physics (for example, we use the Breit-Wigner formula convoluted with a Gaussian function to describe the shape of a resonance). This procedure is common in model-dependent measurements as well as many model-independent measurements. If the signal shape is determined by a model, we usually need to consider theoretical uncertainty due to this model. 
Only one systematical uncertainty source is introduced in the current model. The uncertainty allows the background distribution to deviate from the prediction and the deviation at each bin should behave in a coherent way. Thus we introduce one nuisance parameter $\alpha$. 
However, the MC statistical uncertainty should be considered differently. It is due to limited sample size in the MC simulation. It is usually true that the simulation is done randomly in all bins and there is no bin-by-bin correlation. Hence we introduce one nuisance parameter $\gamma_i$ ($i=1,2,\ldots,N$) for each bin.

Ignoring the constant terms, the log likelihood function is 
\begin{equation}
   \ln\mL = -\frac{\alpha^2}{2} + \sum_{i=1}^N \left[ n_i\ln(\mu s_i+\gamma_i b_i +\alpha\Delta_i) - (\mu s_i +\gamma_i b_i +\alpha\Delta_i) + m_i\ln\gamma_i - \gamma_i m_i\right] \: .
\end{equation}
Unlike previous model considering a single number-counting experiment, the present model can be seen as a combination of $N$ number-counting measurements. It is nearly impossible to solve out the the best-fit $\mu$ analytically by maximizing the present log likelihood function. Let us use an Asimov dataset~\cite{asimov}, where $n_i = b_i + s_i$ for all bins. This option will not change the conclusion in this paper as we are studying the uncertainty of the fitting parameters. The best-fit values are then
\begin{equation}
   \hatmu = 1 \:, \hata = 0\:, \hatg = 1 \: . 
\end{equation}

Letting the parameters are arranged in the order of $\mu,\alpha,\gamma_1,\gamma_2,\cdots,\gamma_N$, the inverse of the covariance matrix $\bV$ is 
\begin{equation}
   \bV^{-1} =
   \begin{pmatrix}
	s\otimes s & s\otimes \Delta & s_1*b_{1} & s_2*b_{2} & \cdots & s_N*b_{N} \\
	s\otimes \Delta & 1 + \Delta\otimes \Delta & \Delta_{1}*b_{1} & \Delta_{2}*b_{2} & \cdots & \Delta_{N}*b_{N} \\
	s_1*b_{1} & \Delta_{1}*b_{1} & m_1 + \frac{b_{1}^2}{n_1} & 0  & \cdots & 0 \\
	s_2*b_{2} & \Delta_{2}*b_{2} & 0 & m_2 + \frac{b_{2}^2}{n_2}  & \cdots & 0 \\
	\vdots & \vdots& \vdots & \vdots & \ddots & \vdots \\
	s_N*b_{N} & \Delta_{N}*b_{N} & 0 & 0 & \cdots & m_N + \frac{b_{N}^2}{n_N} \\
   \end{pmatrix} \: .
\end{equation}
To simplify the expression, the sign $*$ is introduced with the definition $x_i*y_i \equiv \frac{x_iy_i}{n_i}$ and the sign $\otimes$ is introduced with the definition $x\otimes y \equiv \sum_{i=1}^N \frac{x_iy_i}{\sqrt{n_i}^2}$ where the summation is over all bins. Here we keep the form $\sqrt{n_i}^2$ to remind us that it represents the Poisson fluctuation. Let us analyse some matrix elements in the first place.
\begin{itemize}
   \item $s\otimes s = \sum_i \frac{s_i^2}{\sqrt{n_i}^2}$ represents the signal significance compared to the statistical fluctuation indicated by the denominator $\sqrt{n_i}$. We expect that this term determines the measurement precision of the signal strength $\mu$ if no systematic uncertainties or MC statistical uncertainty is present. 
  \item $\Delta \otimes \Delta = \sum_i \frac{\Delta_i^2}{\sqrt{n_i}^2}$ represents the significance of the systematic uncertainty compared to the Poisson statistical fluctuation. If this term is big, we expect that $\alpha$ could be over-constrained and may have a large impact on the $\mu$ uncertainty.
   \item $s\otimes \Delta = \sum_i \frac{s_i\Delta_i}{\sqrt{n_i}^2}$ represents the correlation of the signal shape and the systematic variation. If this term is big, it means that the systematic variation is similar to the signal shape and we expect that $\alpha$ would have a large impact on the $\mu$ uncertainty.
   \item $s_i*b_i$ ($\Delta_i* b_i$) describes the contribution to the shape correlation between signal (systematic uncertainty) and background from the $i$-th bin. 
\end{itemize}

To obtain the covariance matrix itself, we decompose the inverse matrix into two parts, $\bA$ and $\bB$. 
\begin{equation}
   \bV^{-1} = \bold{A}+ \bold{B} = \bA (\bI + \bA^{-1}\bB)
\end{equation}
with $\bI$ being the identity matrix.

The matrix $\bA$ contains all the diagonal elements and the correlation term $s\otimes \Delta$ while the matrix $\bB$ contains all other non-diagonal elements. It should be mentioned that it is easy to calculate the inverse matrix of $\bA$. 
\begin{equation}
   \bold{A} =
   \begin{pmatrix}
	s\otimes s & s\otimes \Delta & 0 & 0 & \cdots & 0 \\
	s\otimes \Delta & 1 + \Delta\otimes \Delta & 0& 0& \cdots & \\
	0 & 0 & m_1 + \frac{b_{1}^2}{n_1} & 0  & \cdots & 0 \\
	0 & 0 & 0 & m_2 + \frac{b_{2}^2}{n_2}  & \cdots & 0 \\
	\vdots & \vdots& \vdots & \vdots & \ddots & \vdots \\
	0 & 0 & 0 & 0 & \cdots & m_N + \frac{b_{N}^2}{n_N} \\
   \end{pmatrix}
\end{equation}
\begin{equation}
   \bold{B} =
   \begin{pmatrix}
	0 & 0 & s_1*b_{1} & s_2*b_{2} & ... & s_N*b_{N} \\
	0 & 0 & \Delta_{1}*b_{1} & \Delta_{2}*b_{2} & ... & \Delta_{N}*b_{N} \\
	s_1*b_{1} & \Delta_{1}*b_{1} & 0 & 0  & ... & 0 \\
	s_2*b_{2} & \Delta_{2}*b_{2} & 0 & 0  & ... & 0 \\
	\vdots & \vdots& \vdots & \vdots & \ddots & \vdots \\
	s_N*b_{N} & \Delta_{N}*b_{N} & 0 & 0 & ... & 0 \\
   \end{pmatrix}
\end{equation}
Therefore the covariance matrix can be expressed in the following way.
\begin{equation}\label{eq:VBA}
   \bV = (\bI + \bA^{-1}\bB)^{-1}\bA^{-1} = (\bI+\sum_{i=1}^{\infty}(-\bA^{-1}\bB)^i)\bA^{-1}
\end{equation}
where the fact that $\bI=(\bI+\bx)(\bI-\bx+\bx^2-\bx^3+\cdots)$ is used. In the Appendix~\ref{app:Identity}, we show that it is valid to apply this identity in our case with the assumptions which will be introduced later. To further simplify the expression, we introduce the following sub-matrices according to the vanishing blocks in $\bV$. 
\begin{equation}
   \bold{V}=
   \begin{pmatrix}
	\bV_1 & \bV_2 \\
	\bV_2^T & \bV_3 \\
   \end{pmatrix}
   \:, \quad
   \bold{A}^{-1} =
   \begin{pmatrix}
	\ba_1 & 0 \\
	0 & \ba_2 \\
   \end{pmatrix}
   \:, \quad
   \bold{B} =
   \begin{pmatrix}
	0 & \bb \\
	\bb^{T} & 0 \\
   \end{pmatrix}
\end{equation}
Here $\ba_1$, $\ba_2$ and $\bb$ are 
\begin{equation}\label{eq:a1}
   \ba_1 = \frac{1}{(1+\Delta\otimes \Delta)s\otimes s-(s\otimes \Delta)^2}
   \begin{pmatrix}
	1 + \Delta\otimes \Delta & - s\otimes \Delta \\
	-s\otimes \Delta & s\otimes s\\
   \end{pmatrix}\:,
\end{equation}
\begin{equation}
   \ba_2 = 
   \begin{pmatrix}
	(m_1+\frac{b_1^2}{n_1})^{-1} & 0 & \cdots & 0 \\
	0 & (m_2+\frac{b_2^2}{n_2})^{-1} & \cdots & 0 \\
	\vdots & \vdots & \ddots & \vdots \\
       0 & 0 & \cdots & (m_N+\frac{b_N^2}{n_N})^{-1} \\
   \end{pmatrix}\:,
\end{equation}
and
\begin{equation}
   \bb = 
   \begin{pmatrix}
	s_1 * b_1 & s_2 * b_2 & \cdots & s_N * b_N \\
	\Delta_1 * b_1 & \Delta_2 * b_2 & \cdots & \Delta_N * b_N \\
   \end{pmatrix}\:.
\end{equation}
Then $\bV_1$ is the covariance matrix for $\mu$ and $\alpha$. From Eq.~\ref{eq:VBA} and letting $\bX\equiv \ba_1\bb\ba_2\bb^T $, we can show that
\begin{equation}\label{eq:V1}
   \bV_1 = (\bI + \bX + \bX^2 + \bX^3 + \cdots) \ba_1 \:.
\end{equation}
To make approximation, we write down the explict expression for $\bX$.

\begin{equation}\label{eq:X}
   \bX = \ba_1
   \begin{pmatrix}
	s\hat{\otimes}s & s\hat{\otimes} \Delta \\
	s\hat{\otimes}\Delta & \Delta\hat{\otimes} \Delta \\
   \end{pmatrix}
   \approx
   \begin{pmatrix}
	\frac{s\hat{\otimes}s}{s\otimes s} & 0\\
	0 & \frac{\Delta\hat{\otimes} \Delta}{1 + \Delta\otimes \Delta} \\
   \end{pmatrix}
\end{equation}
Here the sign $\hat{\otimes}$ is defined as $x\hat{\otimes}y \equiv \sum_{i=1}^N \frac{x_iy_i}{n_i (1 + \frac{n_i}{\delta_i^2})}$. In Eq.~\ref{eq:X}, $\bX$ is made diagonal if we only care about the diagonal elements of $\bV_1$ and  assume that the MC statistical uncertainty have the similar size relative to the Poisson fluctuation in all bins and that the non-diagonal elements are much smaller than the diagonal elements, namely, $(s\otimes \Delta)^2<<(s\otimes s)(\Delta\otimes\Delta)$. The details are presented in the Appendix~\ref{app:X}. These assumptions also allow us to express the covariance matrix as a power series as shown in the Appendix~\ref{app:Identity}. $s\otimes\Delta$ describes the correlation between signal shape and the systematical variation. Thus the approximation is reasonable because the signal shape is usually peaky while the systematic variation is random, which will be estimated quantitatively later. We emphasize that the approximation is only applied to $\bX$, not to $\ba_1$ in Eq.~\ref{eq:V1}. Therefore, this correlation is still partly considered. The diagnonal elements of $\bV_1$ give the uncertainty of $\mu$ and $\alpha$. Defining $\check{\otimes}$ as $x\check{\otimes}y \equiv x\otimes y - x\hat\otimes y=\sum_i \frac{x_iy_i}{n_i + \delta_i^2}$, we have
\begin{equation}\label{eq:dmu}
   \hatdmu =\sqrt{(\bV_1)_{11}} \approx \frac{1}{\sqrt{s\check{\otimes}s \left[1 - \frac{(s\otimes \Delta)^2}{s\otimes s(1+\Delta\otimes \Delta)}\right]}} \: ,
\end{equation}
and
\begin{equation}\label{eq:dalpha}
   \hatda = \sqrt{(\bV_1)_{22}} \approx \frac{1}{\sqrt{(1+\Delta\check{\otimes} \Delta)\left[1 -\frac{(s\otimes \Delta)^2}{s\otimes s(1+\Delta\otimes \Delta)}\right]}} \: .
\end{equation}

From Eq.~\ref{eq:dalpha}, it turns out that the nuisance parameter could be over-constrained if the systematic variation is significant compared to the combination of the Poisson statistical fluctuation and the MC statistical uncertainty as indicated by the term $\Delta \check{\otimes} \Delta=\sum_{i=1}^N\frac{\Delta_i^2}{\sqrt{n_i}^2 + \delta_i^2}$. We also see that the correlation term $s\otimes\Delta$ exists in both Eq.~\ref{eq:dmu} and Eq.~\ref{eq:dalpha}. This term is assumed to be much smaller than $(s\otimes s)(\Delta\otimes \Delta)$ in the derivation. Using the Cauchy-Schwarz inequality, it is true that $(s\otimes\Delta)^2\leq(s\otimes s)(\Delta\otimes\Delta)$. For a quantitative understanding, let us consider a simplified case where the signal events fall in a single bin, the background distribution is uniform and the systematic variation has the same absolute size in all bins. It is easy to show that
\begin{equation}\label{eq:cauchy}
\frac{(s\otimes\Delta)^2}{(s\otimes s)(\Delta\otimes\Delta)} = \frac{(\frac{s\Delta}{n/N})^2}{\frac{s^2}{n/N}\sum_{i=1}^N\frac{\Delta^2}{n/N}} = \frac{1}{N} \: .
\end{equation}
We can see that this assumption is valid as long as the signal shape is peaky enough (as indicated by the factor of $1/N$). The correlation term is of small contribution to the constraint of $\alpha$  in Eq.~\ref{eq:dalpha}. The impact on $\mu$ uncertainty, however, is determined by this correlation as shown in Eq.~\ref{eq:dmu}. 

In the model above, only one systematic source is considered. It is not difficult to extend it to the case of multiple systematic sources. The log likelihood function is  
\begin{equation}\label{eq:logL_Nsys}
   \ln\mL = -\sum_{j=1}^M\frac{\alpha_j^2}{2} + \sum_{i=1}^N\left[ n_i\ln(\mu s_i+\gamma_i b_i +\sum_{j=1}^{M}\alpha_j\Delta_i^{j}) - (\mu s_i +\gamma_i b_i +\sum_{j=1}^M\alpha_j\Delta_i^j) + m_i\ln\gamma_i - \gamma_i m_i\right] \: ,
\end{equation}
where $M$ is the number of systematic items with $M$ nuisance parameters $\alpha_1,\alpha_2,\cdots,\alpha_M$. With the parameters arranged in the order $(\mu,\alpha_1,\alpha_2,\cdots,\alpha_M,\gamma_1,\gamma_2,\cdots,\gamma_N)$, the inverse of the covariance matrix is
\begin{equation} \label{eq:invV_Nsys}
   \bV^{-1} =
   \begin{pmatrix}
	s\otimes s & s\otimes \Delta^1 & s\otimes \Delta^2 & \cdots & s\otimes\Delta^M & s_1*b_{1} & s_2*b_{2} & \cdots & s_N*b_{N} \\
	s\otimes \Delta^1 & 1 + \Delta^1\otimes \Delta^1 & \Delta^1\otimes\Delta^2 & \cdots & \Delta^1\otimes\Delta^M & \Delta_{1}^1*b_{1} & \Delta_{2}^1*b_{2} & \cdots & \Delta_{N}^1*b_{N} \\
	s\otimes\Delta^2 & \Delta^1\otimes\Delta^2 & 1 + \Delta^2\otimes\Delta^2 & \cdots &\Delta^2\otimes\Delta^M & \Delta_1^2*b_1 & \Delta_2^2*b_2&\cdots & \Delta_N^2 * b_N \\
	\vdots & \vdots& \vdots  & \cdots & \vdots & \vdots & \vdots &\cdots & \vdots\\
	s\otimes\Delta^M & \Delta^1\otimes\Delta^M & \Delta^2\otimes\Delta^M & \cdots & 1+\Delta^M\otimes\Delta^M & \Delta_1^M*b_1 & \Delta_1^M*b_2 & \cdots & \Delta_N^M*b_N \\
	s_1*b_{1} & \Delta_{1}^1*b_{1} & \Delta_1^2*b_1 & \cdots & \Delta_1^M*b_1 & m_1 + \frac{b_{1}^2}{n_1} & 0  & \cdots & 0 \\
	s_2*b_{2} & \Delta_{2}^1*b_{2} & \Delta_2^2*b_2 & \cdots & \Delta_2^M*b_2 & 0 & m_2 + \frac{b_{2}^2}{n_2}  & \cdots & 0 \\
	\vdots & \vdots& \vdots & \cdots & \vdots & \vdots & \vdots & \ddots & \vdots \\
	s_N*b_{N} & \Delta_{N}^1*b_{N} &\Delta_N^2*b_N & \cdots & \Delta_N^M*b_N & 0 & 0 & \cdots & m_N + \frac{b_{N}^2}{n_N} \\
   \end{pmatrix} \: .
\end{equation}

We can see that new elements $\Delta^i\otimes\Delta^j = \sum_{k=1}^N \frac{\Delta_k^i\Delta_k^j}{\sqrt{n_k}^2}$ appear. They represent the correlation between two different systematic uncertainties for $i\neq j$.  We further assume that this kind of correlation is small ($(\Delta^i\otimes\Delta^j)^2<<(\Delta^i\otimes\Delta^i)(\Delta^j\otimes\Delta^j)$) and MC statistical uncertainty is also small. It is not difficult to derive (the calculation details and some discussions on the approximation precision can be found in the Appendix~\ref{app:main}) that
\begin{equation}\label{eq:dmu_Nsys}
   \hatdmu \approx \frac{1}{\sqrt{s\otimes s}} \sqrt{1 + \frac{s\hat{\otimes} s}{s\check{\otimes}s} + \sum_{j=1}^M \left[\frac{(s\otimes \Delta^j)^2}{s\otimes s(1+\Delta^j\otimes \Delta^j)} - \sum_{i\neq j} \frac{(s\otimes\Delta^i)(s\otimes\Delta^j)(\Delta^i\otimes\Delta^j)}{s\otimes s (1+\Delta^i\otimes\Delta^i)(1+\Delta^j\otimes\Delta^j)} \right]} \: ,
\end{equation}
and
\begin{equation}\label{eq:dalpha_Nsys}
   \hat{\sigma}_{\alpha_i} \approx \frac{1}{\sqrt{1+\Delta^i \otimes \Delta^i}}\sqrt{1 + \frac{\Delta^i\hat{\otimes}\Delta^i}{1+\Delta^i\check{\otimes}\Delta^i} +\frac{(s\otimes\Delta^i)^2}{s\otimes s(1+\Delta^i\otimes\Delta^i)}+\sum_{j\neq i} \frac{(\Delta^i\otimes\Delta^j)^2}{(1+\Delta^i\otimes\Delta^i)(1+\Delta^j\otimes\Delta^j)}} \: .
\end{equation}
Eq.~\ref{eq:dmu_Nsys} can be decomposed into three terms
\begin{equation}\label{eq:dmu3}
   \hatdmu^2 = \hatdmu^{02} + \sum_{j=1}^M\hatdmu^{\alpha^j 2} + \hatdmu^{\gamma2} \:,
\end{equation}
and the individual terms are
\begin{eqnarray}
   && \hatdmu^{02} = \frac{1}{s\otimes s} \label{eq:impact0}\\
   && \hatdmu^{\alpha^j2}=\hatdmu^{02}\left[\frac{(s\otimes \Delta^j)^2}{s\otimes s(1+\Delta^j\otimes \Delta^j)}-\sum_{i\neq j} \frac{(s\otimes\Delta^i)(s\otimes\Delta^j)(\Delta^i\otimes\Delta^j)}{s\otimes s (1+\Delta^i\otimes\Delta^i)(1+\Delta^j\otimes\Delta^j)}\right] \label{eq:impact_alpha}\\
   && \hatdmu^{\gamma2} = \hatdmu^{02}\frac{s\hat{\otimes} s}{s\check{\otimes}s} \label{eq:impact_gamma}\: .
\end{eqnarray}
They represent the contribution to $\mu$ uncertainty from the data statistics, the systematic uncertainty ($\alpha^j$) and the MC statistical uncertainty ($\gamma$), respectively. 

In many ATLS and CMS measurements involved with fits, a pruning algorithm is usually applied before performing the fit. It is to prune those systematic uncertainties that are not important for the measurement target and thus to reduce the fitting time. In practice, one would always perform two fits with or without using the pruning algorithm in case any important factors are missed. Part of the reason is that most of the pruning conditions originate from intuitive understanding and are not directly related with the signal sensitivity. The formulae above can be used to develop more reliable pruning criteria. 
Here we propose three conditions.
\begin{eqnarray}
 \frac{(s\otimes \Delta)^2}{s\otimes s(1+\Delta\otimes \Delta)} &<& 0.02  \label{eq:prune1}\\
\frac{(s\otimes \Delta)^2}{s\otimes s(1+\Delta\otimes \Delta)} &<& 0.1 \frac{s\hat{\otimes} s}{s\check{\otimes}s} \label{eq:prune2}\\
\frac{s_i^2/[n_i(1 + \frac{n_i}{\delta_i^2})]}{s\check{\otimes}s} &<& 0.02 \label{eq:prune3}
\end{eqnarray}
Basically, we can ignore a systematical uncertainty if its impact on the signal strength uncertainty is much smaller than the impact of data statistics (inequality~\ref{eq:prune1}) or MC statistical uncertainty (inequality~\ref{eq:prune2}). Numerically, we choose 0.02 as the threshold in the inequality~\ref{eq:prune1} because the change of $\hatdmu$ due to omitting the systematical uncertainty is about 1~\% ($\sqrt{1+0.02}\approx 1 +0.01$) and 0.1 in the inequality~\ref{eq:prune1} so that the systematical effect is one order of magnitude smaller than that of the MC statistical uncertainty, but the thresholds can be re-optimized. The effect of the correlation between different systematical uncertainties is usually minor and thus not included in the inequalities (but we can always use the full expression above instead). Similarly, we can also ignore the nuisance parameter corresponding to the MC statistical uncertainty in the $i$-th bin according to the inequality~\ref{eq:prune3}. 
  
 In the end of this section, let us comment on the validity of the formulae.  The main assumption is small correlation between signal shape and systematical variations and small correlation between different systematic uncertainties. The former part is usually true in the searches for resonance-like signals as shown in Eq.~\ref{eq:cauchy}, where the signal shape is peaky while the systematical variation is relatively smooth. However, the latter part is not always true. Taking the top-quark pair ($t\bar{t}$) background in any typical measurement at LHC as example, the $t\bar{t}$ production cross section uncertainty would be anti-correlated with the uncertainty of the tagging efficiency of jets originated from beauty hadrons.  Both uncertainties affect the $t\bar{t}$ background normalization and this correlation may be not small inevitably. If both systematical uncertainties happen to be important to the signal sensitivity, we admit that it is not precise to calculate $\hat{\sigma}_\mu$ using Eq.~\ref{eq:dmu_Nsys}. On the other hand, we can always check the precision by looking at the omitted sub-leading terms. For $\hat{\sigma}_{\mu}$, these terms (see Appendix~\ref{app:main}) look like
\begin{eqnarray}
 && \sum_{i\neq j}\frac{(s\otimes\Delta^i)^2}{s\otimes s(1+\Delta^i\otimes\Delta^i)}\frac{(s\otimes\Delta^j)^2}{s\otimes s(1+\Delta^j\otimes\Delta^j)} \nonumber \\
+ && \sum_{i\neq j, i\neq k, j\neq k} \frac{(s\otimes\Delta^i) (\Delta^i\otimes\Delta^j) (\Delta^j\otimes\Delta^k) (s\otimes\Delta^k)}{s\otimes s(1+\Delta^i\otimes\Delta^i)(1+\Delta^j\otimes\Delta^j)(1+\Delta^k\otimes\Delta^k)} 
\end{eqnarray}
They are important only when there are multiple systematical uncertainties that are mutually highly correlated or highly correlated with the signal shape. But we expect that this case is not often seen under normal circumstances.  The other assumption is small MC statistical uncertainty. This is usually true. Otherwise, one would probably seek for data-driven methods or request to produce larger MC samples.


\section{\label{sec:example} A pseudo experiment for illustration}

\begin{figure}
   \includegraphics[width=0.5\textwidth]{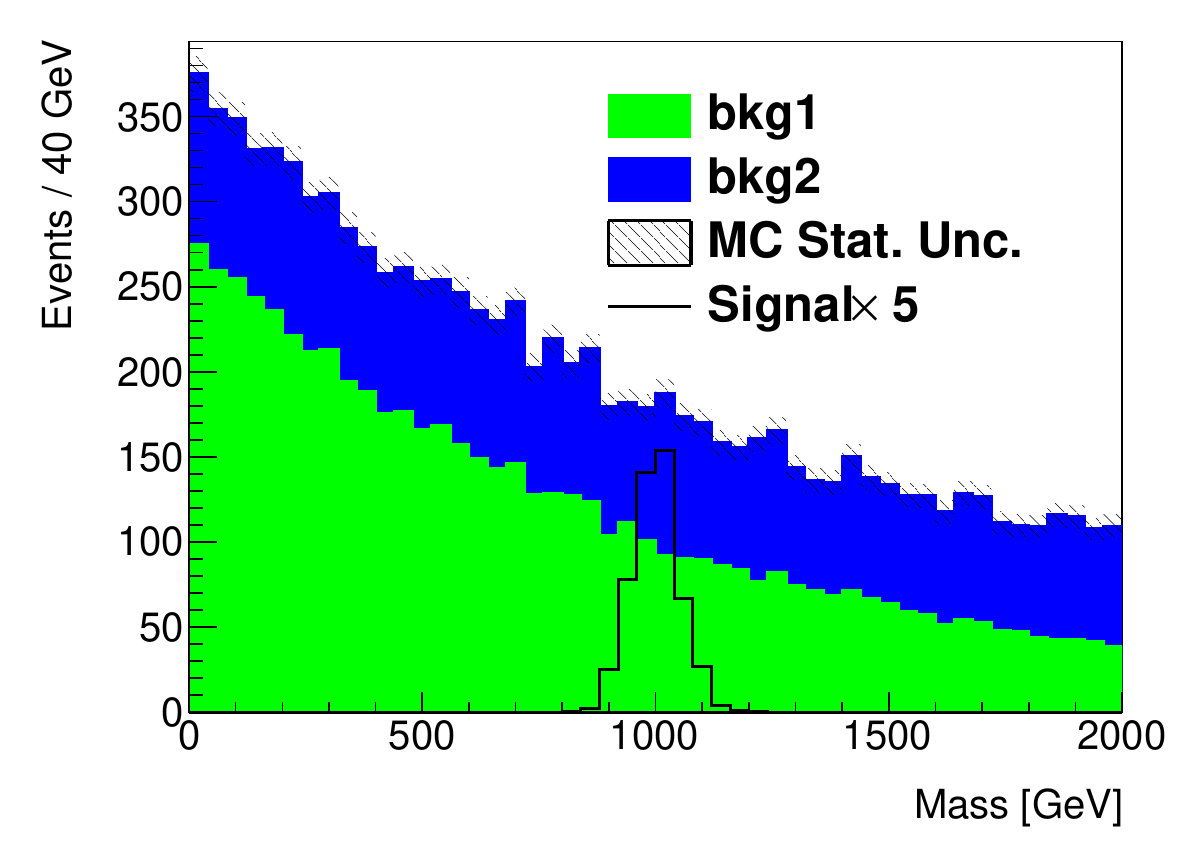}
   \caption{\label{fig:exam_nom}
   The signal and background distributions in the presudo experiment and the signal is scaled by a factor of 5 for illustration.
   }
\end{figure}

In this section, we present a pseudo experiment of searching for a resonance on a mass spectrum. The signal is simulated with a Gaussian distribution $G(x|1000,50)$ with the resonance mass 1000~GeV and mass resolution 50~GeV. We assume the width of the resonance is small compared to the mass resolution. Two background components (denoted by ``bkg1'' and ``bkg2'') are introduced and simulated with exponential distributions with the decay parameter 1000~GeV and 5000~GeV. Each signal/background event is given a constant weight for simplicity. Table~\ref{tab:example_model} summarizes the information for the signal and background models. They are also shown in Fig.~\ref{fig:exam_nom}. 

\begin{table}
   \caption{\label{tab:example_model} Information for the signal and background models.
   }
   \begin{ruledtabular}
	\begin{tabular}{l l l l}
	   & Weighted number of events & Model & Event wegiht\\
	   \hline
	   Signal & 100 & Gauss distribution $G(x|1000, 50)$ & 0.1 \\
	   bkg1 & 6000 & Exponential distribution $e^{-x/1000}$ & 0.2 \\
	   bkg2 & 4000 & Exponential distribution $e^{-x/5000}$ & 0.5 \\
	\end{tabular}
   \end{ruledtabular}
\end{table}

\begin{figure}
   \includegraphics[width=0.32\textwidth]{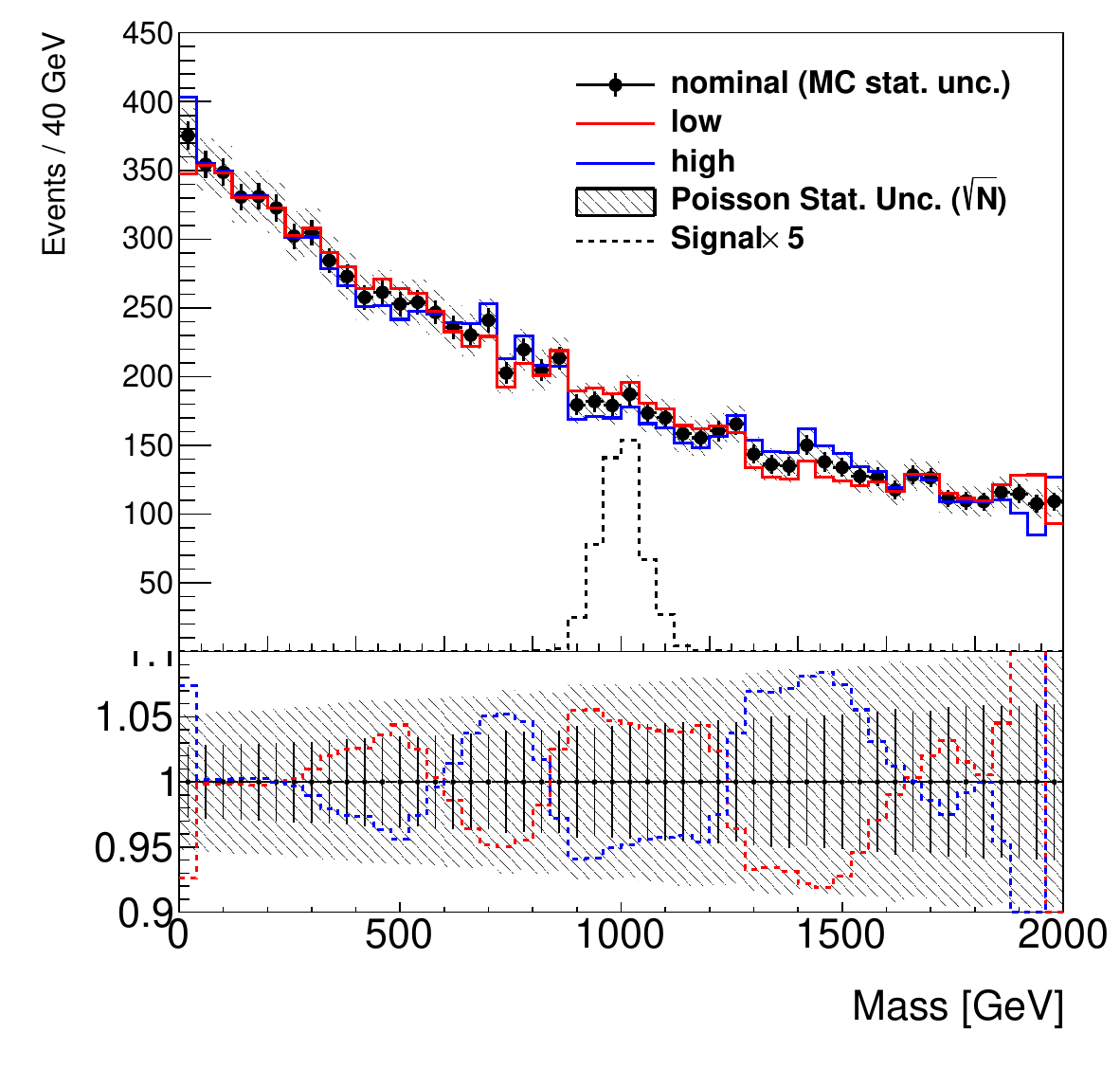}
   \includegraphics[width=0.32\textwidth]{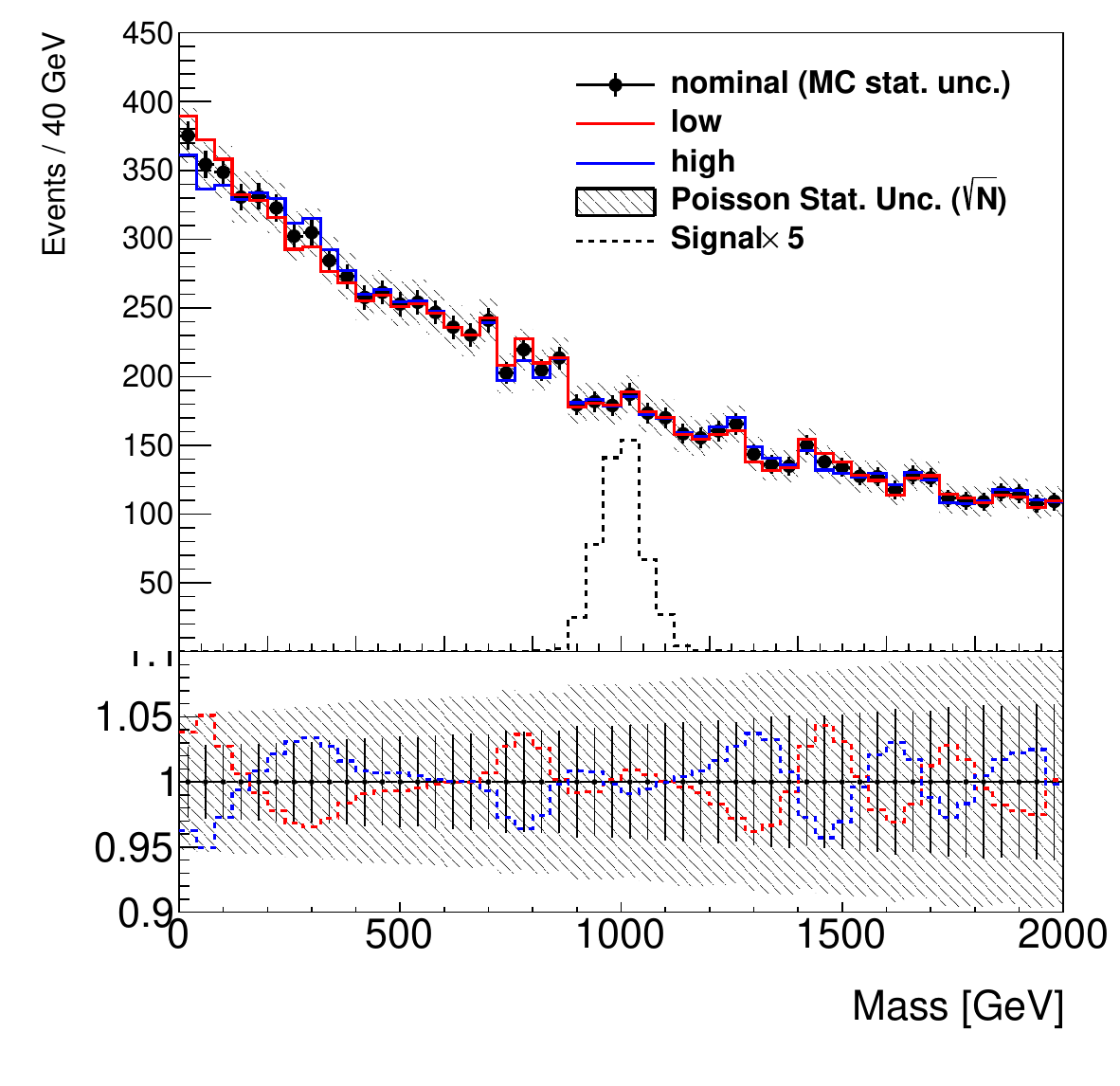}
   \includegraphics[width=0.32\textwidth]{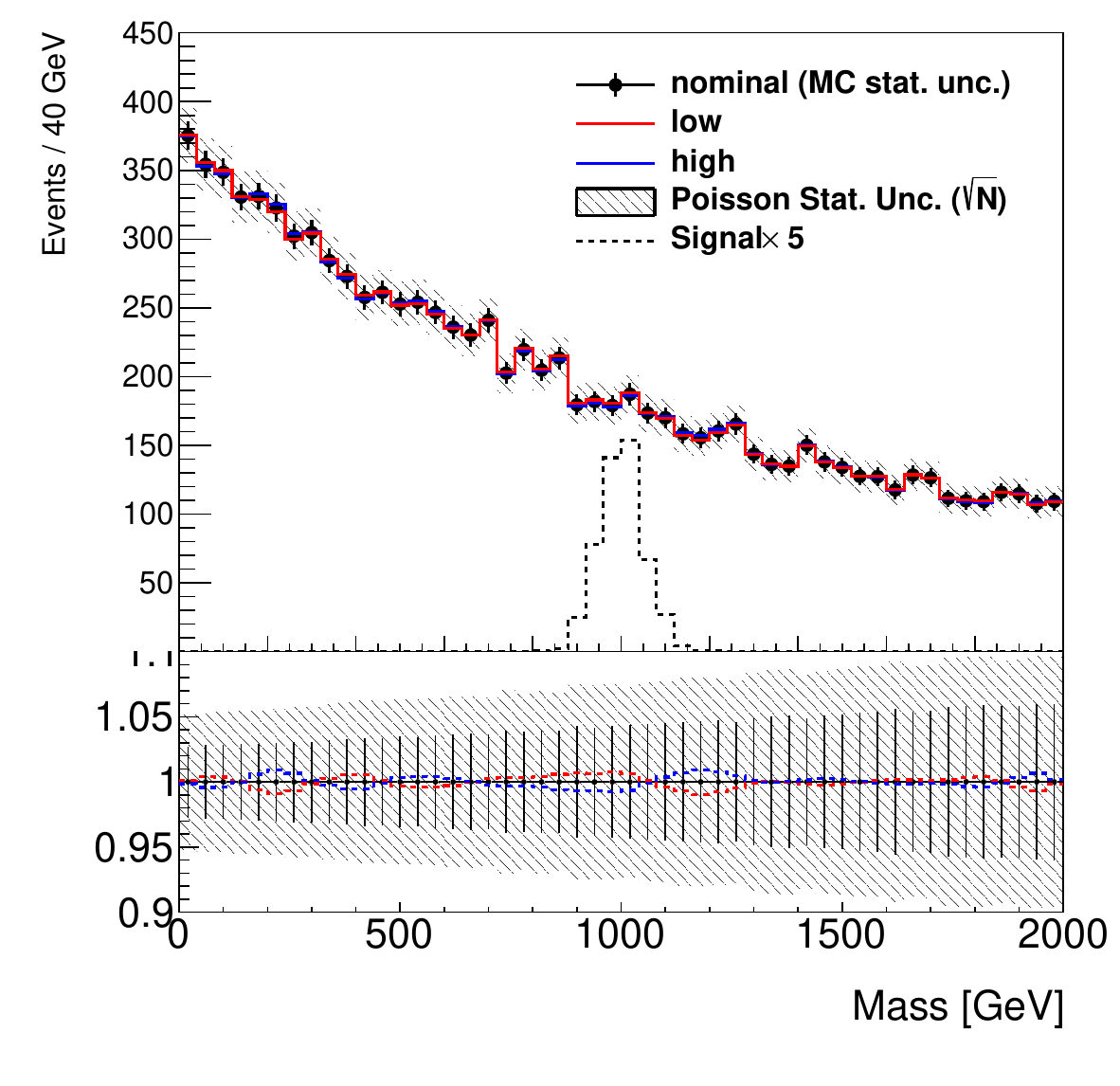}
   \caption{\label{fig:exam_envelope1}
	Envelope plots for the shape-only systematic uncertainties, namely, ShapeBig (L), ShapeMedium (M) and ShapeSmall (R). The blue/red histograms represent the high/low variation of the systematic item. The signal is scaled by a factor of 5 for illustration. The lower pad shows the ratio of systematic variation and the nominal distribution. The black vertical error bars represent the MC statistical uncertainty and the hatch histogram represents the Poisson statistical fluctuation. 
   }
\end{figure}

\begin{figure}
   \includegraphics[width=0.45\textwidth]{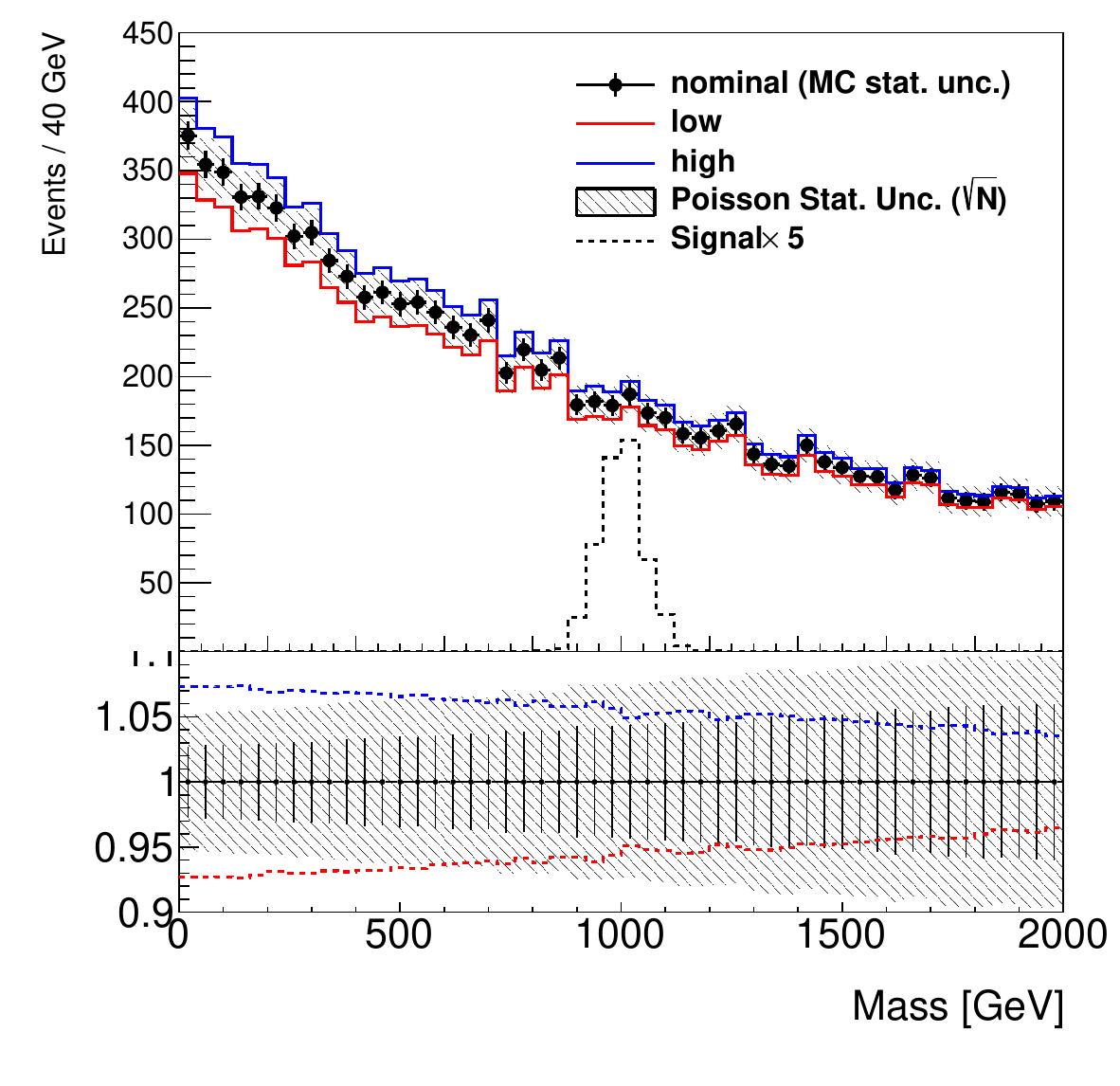}
   \includegraphics[width=0.45\textwidth]{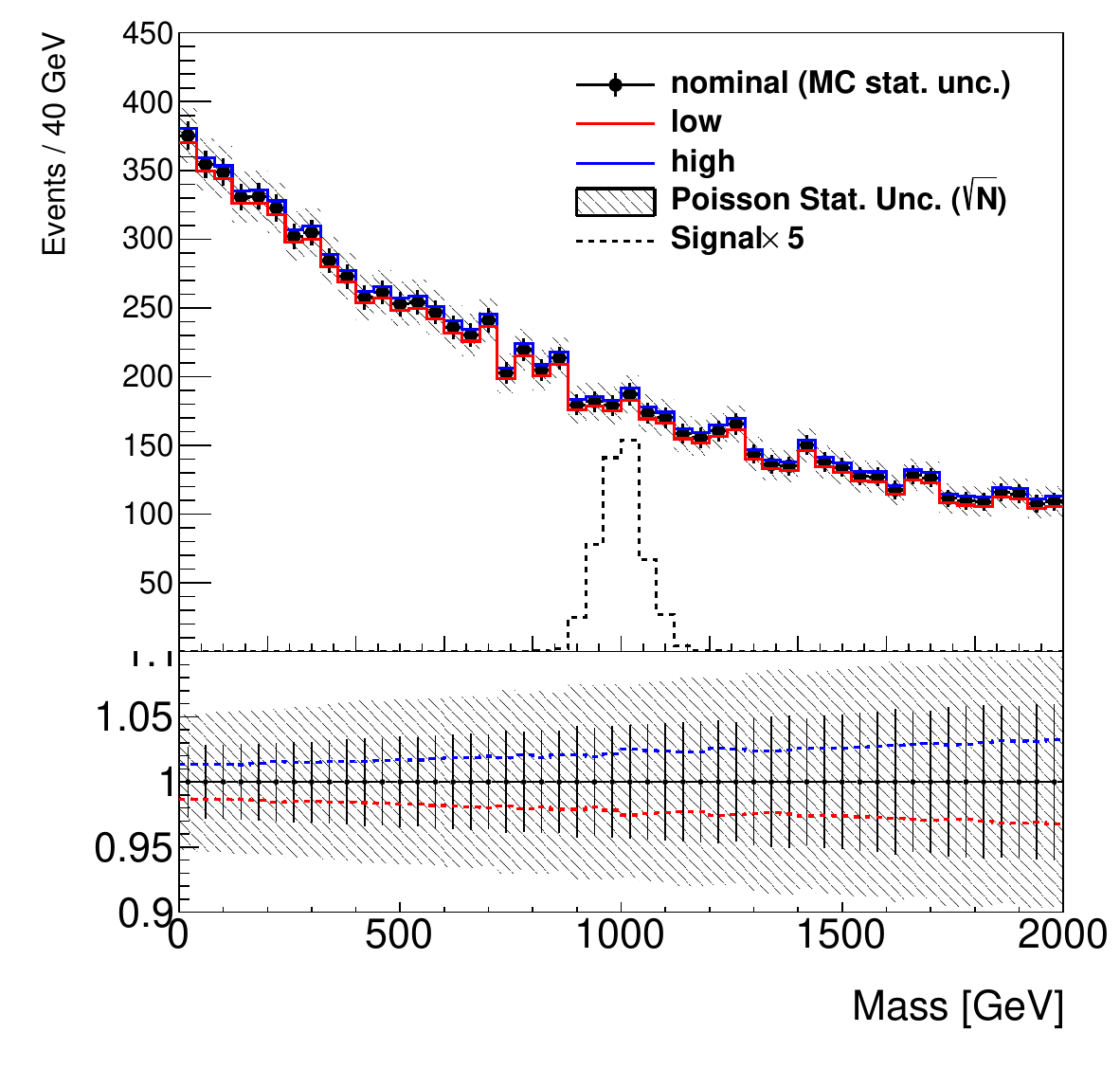}
   \caption{\label{fig:exam_envelope2}
	Envelope plots for the norm-only systematic uncertainties, namely, NormBig (L) and NormSmall (R). The blue/red histograms represent the high/low variation of the systematic item. The signal is scaled by a factor of 5 for illustration. The lower pad shows the ratio of systematic variation and the nominal distribution. The black vertical error bars represent the MC statistical uncertainty and the hatch histogram represents the Poisson statistical fluctuation. 
   }
\end{figure}

We introduce 5 nuisance parameters, where 3 NPs represent 3 shape-only systematic uncertainties (applied to both signal and background components) and 2 NPs represent 2 norm-only systematic uncertainties (applied to bkg1 and bkg2 separately). Here ``shape-only'' means that the systematic item only affects the shape of the observable distribution while ``norm-only'' means it only affects the normalization. To implement the shape-only systematic uncertainties, we apply a random variation to the nominal distribution in each bin, where the random variation is generated with a Gaussian distribution with the mean 0 and different standard deviations. 
As shown in Table~\ref{tab:example_systs},  ``ShapeBig/ShapeMedium/ShapeSmall'' label the size of shape-only the systematic effect. They correpsond to a standard deviation of $\Delta n$, $0.5\Delta n$, and $0.1\Delta n$ respectively. Here $\Delta n$ denotes the statistic uncertainty in that bin. A normalization uncertainty of 10~\% is applied to bkg1 and denoted by ``NormBig'' while that of 5~\% is applied to bkg2 and denoted by ``NormSmall''. 
All the systematic uncertainties are summarized in Table~\ref{tab:example_systs}. The envelope plots for the systematic uncertainties are shown in Fig.~\ref{fig:exam_envelope1} and Fig.~\ref{fig:exam_envelope2}. In these plots, the blue/red histogram represents the ``high''/``low'' variation for a systematic item. They represent the initial estimation of this systematic uncertainty. In addition, MC statistical uncertainty is also considered.

\begin{table}
   \caption{\label{tab:example_systs} Information for the systematic uncertainties. Here $\Delta n$ denotes the total statistical uncertainty. 
   }
   \begin{ruledtabular}
	\begin{tabular}{l l l}
	   Nuisance parameter name  & Applied to sample & Variation Size \\
	   \hline
	   ShapeBig & signal, bkg1, bkg2 &$G(x|0,\Delta n)$ \\
	   ShapeMedium &signal, bkg1, bkg2 & $G(x|0, 0.5\Delta n)$ \\
	   ShapeSmall & signal, bkg1, bkg2 & $G(x|0, 0.1\Delta n)$ \\
	   NormBig &  bkg1 & 10\% \\
	   NormSmall &bkg2 &  5\%  \\
	\end{tabular}
   \end{ruledtabular}
\end{table}

For this toy measurement, the fit is performed using a tool based on the HistoFactory~\cite{histofactory}, where advanced numerical tools are used to determine the covariance matrix precisely. In Table~\ref{tab:example_results}, the fitting results and the approximate calculations using the Eqs.~\ref{eq:dalpha_Nsys}, \ref{eq:impact0}, \ref{eq:impact_alpha} and \ref{eq:impact_gamma} are summarized for comparison. To use the equations, we have (taking $s\otimes \Delta$ as example)
\begin{equation}\label{eq:example_SB}
   s\otimes\Delta^{\text{high/low}} = \sum_{i=1}^N \frac{S_i(N_i^{\text{high/low}}-N_i^{\text{nom}})}{\sqrt{N_i^{\text{nom}}}^2} \:  ,
\end{equation}
where $S_i$ is the number of signal events in $i$-th bin; $N_i^{\text{nom}}$ is the predicted total number of events; $N_i^{\text{high/low}}$ is the total number of events corresponding to the systematic ``high/low'' variation.

We can see that only the ShapeSmall NP is mildly constrained while all others are over-constrained. The calculated constraint is consistent with that from the fit for the shape-only systematic uncertainties while this consistence is not very good for the norm-only systematic uncertainties. One of the reasons is that we are using a linear interpolation strategy instead of the exponential interpolation strategy used in the fit~\cite{histofactory} when implementing the norm-only systematic uncertainty. Taking the model in Sec.~\ref{sec:simple_model} as example, the linear interpolation is $b\pm\alpha\Delta$ while the exponential interpolation is $b(1\pm\frac{\Delta}{b})^{\alpha^\prime}$ (a ``$\prime$'' is added to distinguish from that in linear interpolation). $\alpha\approx\alpha^\prime$ only if $\Delta/b$ is small and we can show that $\sigma_\alpha = \frac{\ln(1+\Delta/b)}{\Delta/b}\sigma_{\alpha^\prime}<\sigma_{\alpha^\prime}$. The calculated impact is also fairly consistent with that in the fit although the correlation between different systematic items and the correlation between signal shape and the systematic variation are not fully considered. Especially, it should be noted that the calculation is able to indicate which systematic items would be important. For example, ShapeMedium would have a larger impact than ShapeSmall as its size is bigger by our design. But from either the fit or the approximate calculation, its impact is smaller. There are two reasons behind. One is that the ShapeMedium variation is larger and thus the corresponding NP is more constrained. The other is that the correlation between the signal shape and the ShapeSmall variation turns out to be bigger than the ShapeMedium variation. The latter point can be also seen by comparing the middle and right plots in Fig.~\ref{fig:exam_envelope1}. Similarly, we find that the NormSmall systematic item has actually a larger impact than the NormBig systematic item. 

\begin{table}
   \caption{\label{tab:example_results} 
   Comparison of the fitting results and the approximate calculations.
   }
   \begin{ruledtabular}
	\begin{tabular}{l | l l  | l l }
	   & \multicolumn{2}{l|}{Constraint ($\hat{\sigma}_{\alpha}$ in Eq.~\ref{eq:dalpha_Nsys})}  & \multicolumn{2}{l}{Impact ($\hatdmu^0$,$\hatdmu^{\alpha}$ and $\hatdmu^{\gamma}$ in Eqs.~\ref{eq:impact0}-\ref{eq:impact_gamma})} \\
	   \hline
	   Nuisance parameter name & Fit & Calculation & Fit & Calculation\\
	   \hline
	   ShapeBig & $\pm0.26$  & $\pm0.24$ & $ _{-0.116}^{+0.112}$ & $_{-0.080}^{+0.082}$\\
	   ShapeMedium & $\pm0.48$ & $\pm0.48$ & $_{-0.045}^{+0.038}$ & $_{-0.015}^{+0.015}$\\
	   ShapeSmall & $\pm0.94$ & $\pm0.94$ & $_{-0.045}^{+0.045}$ & $_{-0.040}^{+0.040}$\\
	   NormBig &  $\pm0.29$ & $\pm 0.23$ &$_{-0.015}^{+0.024}$& $_{-0.022}^{+0.022}$\\
	   NormSmall & $\pm0.78$ & $\pm 0.61$ &$_{-0.058}^{+ 0.051}$& $_{-0.044}^{+0.044}$\\
	   MC Stat. Unc. &  & & $_{-0.19}^{+0.18}$ & $\pm0.17$ \\
	   Data statistics &  & & $_{-0.28}^{+0.29}$ & $\pm0.29$ \\
	\end{tabular}
   \end{ruledtabular}
\end{table}

In this pseudo experiment, the norm and shape components of a systematical item are considered separately. But it is trivial to apply the calculation if a systematic item affects both the normalization and shape of the observable distribution. 


\section{\label{sec:summary}Summary}
In summary, the constraint and impact on the POI of nuisance parameters in maximum likelihood method are studied. Based on simplified models, we find that a nuisance parameter will be over-constrained if the corresponding variation is large compared to the total statistical uncertainty (the combination of the Poisson statistic fluctuation and the MC statistic uncertainty). It will have a large impact on the POI uncertainty if the corresponding variation has a strong correlation with the signal shape. Assuming small MC statistical uncertainty, small correlation between different systematic uncertainties and small correlation between signal shape and systematic variation, simple formulae (Eqs.~\ref{eq:dmu_Nsys},~\ref{eq:dalpha_Nsys},~\ref{eq:impact0},~\ref{eq:impact_alpha}, and \ref{eq:impact_gamma}) are derived to calculate the constraint and impact. A toy experiment is also performed and shows fair consistence between the calculation and that using the current fitting tool. In many measurements by ATLAS or CMS collaborations, complicated fits are involved and advanced numerical methods are developed to obtain the covariance matrix very precisely in the fitting tools. This study is not to provide a replacement of the numerical methods, but helps to cross-check the potential features in the fitting results in an easy and direct way. It can also help us to improve the pruning algorithms adopted in many fitting tools.

\section{Acknowledgement}
I admit that this work will not be finished without the close cooperation with Christopher John Mcnicol, Elisabetta Pianori and Paul Thompson. I would also like to thank Liang Zhong and Mengzhen Wang for discussions on some mathematics topics and Fang Dai for encouraging words.

\begin{appendix}
    \section{Validation of the identity of $\bI = (\bI+\bx)(\bI-\bx+\bx^2-\bx^3+\cdots)$}\label{app:Identity}
In this section, let us explain it is valid to apply the identity of $\bI = (\bI+\bx)(\bI-\bx+\bx^2-\bx^3+\cdots)$ in our case. In the first place, it is not difficult to show that the sufficient and necessary condition for this identity is that the absolute value of every eigenvalue of $\bx$ is less than 1. We then show this condition is satisfied in our case with $\bx = \bA^{-1}\bB$. Using the sub-matrices $\ba_1$, $\ba_2$ and $\bb$ defined in Sec.~\ref{sec:realistic_model}, we have
\begin{equation}
    \bx = \begin{pmatrix}
        0 & \ba_1\bb \\
        \ba_2\bb^T & 0 \\
    \end{pmatrix}\: .
\end{equation}
The eigenvalues can be found by solving the equation $\det|\bx-\lambda\bI|=0$. We resort to Schur's determinant identity, namely,
\begin{equation}\label{eq:schur}
    \det\left|\begin{matrix} \bA & \bB \\
    \bC & \bD \\
\end{matrix}\right| = \det|\bD|\det|\bA-\bB\bD^{-1}\bC| 
\end{equation}
which holds if $\bD$ is invertible. Applying it to $\det|\bx-\lambda\bI|$, we have
\begin{equation}
    \det|\bx-\lambda\bI| = \det\left|\begin{matrix}
        -\lambda \bI_{2\times 2} & \ba_1\bb \\
        \ba_2\bb^T & -\lambda \bI_{N\times N} \\
    \end{matrix}\right|
    =\det|-\lambda\bI_{N\times N}|\det|-\lambda\bI_{2\times 2}+\frac{1}{\lambda}\bX| \: ,
\end{equation}
where $\bI_{2\times 2}$ and $\bI_{N\times N}$ denote the $2\times 2$ and $N\times N$ identity matrices respectively and $\bX\equiv \ba_1\bb\ba_2\bb^{T}$ as defined in Sec.~\ref{sec:realistic_model}. Assuming $(s\otimes s)(\Delta\otimes\Delta) >> (s\otimes\Delta)^2$, we find that $\bX$ is approximately an upper triangular matrix as shown in Appendix~\ref{app:X}. Using Eq.~\ref{eq:appX_X}, it is easy to obtain
\begin{equation}
    \det|\bx-\lambda\bI| \approx (-\lambda)^{N}(-\lambda + \frac{\bX_{11}}{\lambda})(-\lambda + \frac{\bX_{22}}{\lambda}) \:.
\end{equation}
Therefore, the only non-vanishing eigenvalues are $|\lambda|=\sqrt{\bX_{11}}\approx \sqrt{\frac{s\hat{\otimes}s}{s\otimes s}}< 1$, and $ |\lambda|=\sqrt{\bX_{22}}\approx \sqrt{\frac{\Delta\hat{\otimes}\Delta}{1+\Delta\otimes\Delta}}<1$. Both of them are less than 1, which guarantee that we can apply the identity $\bI = (\bI+\bx)(\bI-\bx+\bx^2-\bx^3+\cdots)$ with the assumption $(s\otimes s)(\Delta\otimes\Delta) >> (s\otimes\Delta)^2$.

\section{Approximation in the expression of $\bX$}\label{app:X}
In this section, let us explain that the approximation in Eq.~\ref{eq:X} is from the main assumption $\epsilon \equiv \frac{(s\otimes\Delta)^2}{(s\otimes s)(\Delta\otimes\Delta)}<<1$. First of all, we can ignore the difference between $\hat{\otimes}$ and $\otimes$ in Eq.~\ref{eq:X} by assuming the MC statistical uncertainty has the same size relative to the Poisson fluctuation, namely, $\frac{\delta_i}{\sqrt{n_i}} = r$ for all bins. This assumption
is unnecessary because we only need $\frac{(s\otimes\Delta)^2}{(s\otimes s)(\Delta\otimes\Delta)}<<1$ which also leads to $\frac{(s\hat{\otimes}\Delta)^2}{(s\otimes s)(\Delta\otimes\Delta)}<<1$ with $0<s\hat{\otimes}\Delta<s\otimes\Delta$. But we keep using this assumption as it is usually true and brings convenience. Using Eq.~\ref{eq:a1}, Eq.~\ref{eq:X} becomes 


\begin{eqnarray}\label{eq:appX_X}
    \bX &=& \ba_1 
    \begin{pmatrix}
    s\hat{\otimes}s & s\hat{\otimes}\Delta \\
    s\hat{\otimes}\Delta & \Delta\hat{\otimes}\Delta\\
\end{pmatrix} \approx c\ba_1 
\begin{pmatrix}
              s\otimes s & s \otimes\Delta \\
              s\otimes\Delta & \Delta\otimes\Delta\\
\end{pmatrix}\\
\bX_{11} &\approx& c \\
\bX_{12} &\approx& c \frac{s\otimes\Delta}{(1+\Delta\otimes\Delta)s\otimes s - (s\otimes\Delta)^2} \approx c\frac{\epsilon_1}{s\otimes\Delta}\\
\bX_{21} &\approx& 0 \\
\bX_{22} &\approx& c\frac{(s\otimes s)(\Delta\otimes\Delta) - (s\otimes\Delta)^2}{(1+\Delta\otimes\Delta)s\otimes s - (s\otimes\Delta)^2} \approx c\frac{\Delta\otimes\Delta}{1+\Delta\otimes\Delta}(1-\frac{\epsilon}{1+\Delta\otimes\Delta}) \label{eq:X22}
\end{eqnarray}
where $c=\frac{1}{1+\frac{1}{r^2}} \approx \frac{s\hat{\otimes}s}{s\otimes s} \approx \frac{\Delta\hat{\otimes}\Delta}{\Delta\otimes \Delta}$ and $\epsilon_1 \equiv \frac{\Delta\otimes\Delta}{1+\Delta\otimes\Delta}\epsilon < \epsilon$. 
We note that $\bX$ is approximately an upper triangular matrix. 
By the assumption $\epsilon<<1$, we can neglect the term proportional to $\epsilon$ for the element $\bX_{22}$, but we cannot neglect $\bX_{12}$ as it is possible that $s\otimes\Delta$ is small and $\frac{\epsilon_1}{s\otimes\Delta}$ is big. 

The trick is that we can neglect $\bX_{12}$ as long as we only care about the diagonal elements in the covariance matrix, $\bV_1$. Noting that $\bV_1 = (1+\bX + \bX^2+\ldots)\ba_1$, we can show it order by order. The zeroth-order term is $\bV^{(0)}=\ba$ (in this section, we omit the subscript ``1'' in $\bV_1$ and $\ba_1$ for cleaness.) and does not contain terms about $\bX_{12}$. Let us start with the first-order term, $\bV^{(1)}=\bX\ba$ and focus on the diagonal element $\bV_{11}^{(1)}$ (for the
other one $\bV_{22}^{(1)}$, any term containing $\bX_{12}$ will always contain $\bX_{21}$ and thus be vanishing. This is also true for higher-order terms in $\bV$.). 
\begin{eqnarray}\label{eq:v1}
    \bV_{11}^{(1)} = \bX_{11}\ba_{11}(1 + \frac{\bX_{12}\ba_{21}}{\bX_{11}\ba_{11}}) \approx \bX_{11}\ba_{11}(1 - \frac{\epsilon_1}{1+\Delta\otimes\Delta} )
\end{eqnarray}
We can neglect the term proportional to $\epsilon$, which is equivalent to neglecting the element $\bX_{12}$. Let us rewrite the equation above as $\frac{\bV_{11}^{(1)}}{\bX_{11}\ba_{11}}=1+\mathcal{O}(\epsilon)$ where $\mathcal{O}(\epsilon)$ denotes any term of the order of $\epsilon$. For the second-order term $\bV^{(2)}=\bX^2\ba$, the diagonal element can be expressed as (noting that $\bV^{(2)}= \bX\bV^{(1)}$)
\begin{eqnarray}\label{eq:v2}
    \bV_{11}^{(2)}&=& \bX_{11}\bV_{11}^{(1)} ( 1 + \frac{\bX_{12}\bV_{21}^{(1)}}{\bX_{11}\bV_{11}^{(1)}}) \\
                  &=& \bX_{11}\bV_{11}^{(1)} ( 1 + \frac{\bX_{12}\bX_{22}\ba_{21}}{\bX_{11}\bX_{11}\ba_{11}}) \\
                  &=& \bX_{11}\bV_{11}^{(1)} \left[ 1 + \frac{\bX_{22}}{\bX_{11}}\left(\frac{\bV_{11}^{(1)}}{\bX_{11}\ba_{11}}-1\right)\right] \: .
\end{eqnarray}
We can see that $\frac{\bV_{11}^{(2)}}{\bX_{11}\bV_{11}^{(1)}} = 1 + \mathcal{O}(\epsilon)$ because of Eq.~\ref{eq:v1}. Similarly, the third-order term $\bV_{11}^{(3)}$ is
\begin{equation}
    \bV_{11}^{(3)}= \bX_{11}\bV_{11}^{(2)} \left[ 1 + \frac{\bX_{22}}{\bX_{11}}\left(\frac{\bV_{11}^{(2)}}{\bX_{11}\bV_{11}^{(1)}}-1\right)\right] \: ,
\end{equation}
and thus $\frac{\bV_{11}^{(3)}}{\bX_{11}\bV_{11}^{(2)}}=1+\mathcal{O}(\epsilon)$.
Using the method of mathematical induction, we can show that $\frac{\bV_{11}^{(n+1)}}{\bX_{11}\bV_{11}^{(n)}}=1+\mathcal{O}(\epsilon)$ for any $n\geq 1$. Therefore, if we only care about the diagonal elements of $\bV$, neglecting the terms of the order of $\epsilon$ is equivalent to that we neglect $\bX_{12}$ in the beginning and hence $\bX$ is made diagonal in Eq.~\ref{eq:X}. 


\section{Some calculation details}\label{app:main}
In this section, we present some calculation details to derive Eq.~\ref{eq:dmu_Nsys} and Eq.~\ref{eq:dalpha_Nsys} in the case of $M$ systematical uncertainty sources.  The matrices $\bV$, $\bA^{-1}$ and $\bB$ can be written in the following form.
\begin{equation}
   \bold{V}=
   \begin{pmatrix}
	\bV_1 & \bV_2 \\
	\bV_2^T & \bV_3 \\
   \end{pmatrix}
   \:, \quad
\bA^{-1} = \begin{pmatrix} \ba_1 & 0 \\  0 & \ba_2 \end{pmatrix} \:, \quad 
\bB = \begin{pmatrix} \ba_3 & \bb \\ \bb^T & 0 \end{pmatrix} \: ,
\end{equation}
where the sub-matrices $\ba_2$ and $\bb$ are the same as in the main text while $\ba_1$ and $\ba_3$ are 
\begin{eqnarray}
\ba_1 &=& \begin{pmatrix}
 \frac{1}{s\otimes s} &  & & & \\
 &\frac{1}{\Delta^1\otimes\Delta^1} & & & \\
 &&\frac{1}{\Delta^2\otimes\Delta^2} & & \\
 &&& \ddots & \\
 &&&& \frac{1}{\Delta^M\otimes\Delta^M} \\
 \end{pmatrix} \:,\\
 \ba_3 &=& \begin{pmatrix}
 0 & s\otimes\Delta^1 & s\otimes\Delta^2 & \cdots & s\otimes\Delta^M\\
 s\otimes\Delta^1&0 & \Delta^1\otimes\Delta^2 &\cdots & \Delta^1\otimes\Delta^M \\
 s\otimes\Delta^2&\Delta^1\otimes\Delta^2&0&\cdots &\Delta^2\otimes\Delta^M \\
 \vdots&\vdots&\vdots& \ddots & \vdots \\
 s\otimes\Delta^M&\Delta^1\otimes\Delta^M&\Delta^2\otimes\Delta^M&\cdots &0 \\
 \end{pmatrix} \: .
\end{eqnarray}

We assume that the covariance matrix $\bV$ can be expressed as a series, namely, $\bV = [1+\sum_{i=1}^{+\infty}(-\bA^{-1}\bB)^i]\bA^{-1}$.  Let us investigate $\bx \equiv \bA^{-1}\bB$ to check this assumption.
\begin{equation}
\bx = \begin{pmatrix}
\ba_1\ba_3 & \ba_1\bb \\ \ba_2\bb^T & 0 
\end{pmatrix}
\end{equation}
Using the same procedure presented in Appendix~\ref{app:Identity}, the eigenvalues of $\bx$ can be found from the equation below.
\begin{eqnarray}
\det|\bx| &=& \det|-\lambda\bI_{N\times N}|\det|\ba_1\ba_3 - \lambda\bI_{1+M,1+M} + \frac{1}{\lambda}\bX| \label{eq:detx} \\
  &\approx& (-\lambda)^N(-\lambda+\frac{1}{\lambda}\frac{s\hat{\otimes}s}{s\otimes s})\Pi_{i=1}^M\left(-\lambda+\frac{1}{\lambda}\frac{\Delta^i\hat{\otimes}\Delta^i}{1+\Delta^i\otimes\Delta^i}\right) \label{eq:detx1}
\end{eqnarray}
where $\bX\equiv \ba_1\bb\ba_2\bb^T$, the same definition as in the main text. The terms containing $\bX$ arise from the MC statistical uncertainty. It is difficult to solve Eq.~\ref{eq:detx}. But if the correlations are small, namely, $\frac{(s\otimes\Delta^i)^2}{(s\otimes s)(\Delta^i\otimes\Delta^i)}<<1$ and $\frac{(\Delta^i\otimes\Delta^j)^2}{(\Delta^i\otimes\Delta^i)(\Delta^j\otimes\Delta^j)}<<1$, and MC statistical uncertainty is small, we can expect the determinant in Eq.~\ref{eq:detx} is dominated by the contribution from the diagonal elements and hence we have Eq.~\ref{eq:detx1} in the limits $s\otimes\Delta^i \to 0$, $\Delta^i\otimes\Delta^j \to 0$ and $\delta^i \to 0$. Noting that $0<s\hat{\otimes}s<s\otimes s$ and $0<\Delta^i\hat{\otimes}\Delta^i < \Delta^i\otimes\Delta^i$, it is easy to see that the absolute value of all eigenvalues is less than 1 and thus it is valid for this series expansion.

Now let us present some calculation details for the main results Eq.~\ref{eq:dmu_Nsys} and~\ref{eq:dalpha_Nsys}. We are interested in the top left block of the $\bV$, $\bV_1$. The first few terms are
\begin{eqnarray}
\bV_1^{(0)} &=& \ba_1 \: ,\\
\bV_1^{(1)} &=& -(\ba_1\ba_3)\ba_1 \:, \\
\bV_1^{(2)} &=& [(\ba_1\ba_3)^2 + \bX] \ba_1 \:, \\
\bV_1^{(3)} &=&-[(\ba_1\ba_3)^3 + (\ba_1\ba_3)\bX + \bX(\ba_1\ba_3) ] \ba_1 \:,\\
\bV_1^{(4)} &=&[(\ba_1\ba_3)^4 + (\ba_1\ba_3)^2\bX + (\ba_1\ba_3)\bX(\ba_1\ba_3) + \bX(\ba_1\ba_3)^2 + \bX^2 ] \ba_1 \:.
\end{eqnarray}
Let us focus on the diagonal elements of $\bV_1$. We find that 
\begin{eqnarray}
(\ba_1\ba_3\bX)_{ii} &=& \sum_{j\neq i} (\ba_1)_{ii}(\ba_3)_{ij}(\ba_1)_{jj}(\bb\ba_2\bb^T)_{ij} \:, \\
(\bX\ba_1\ba_3)_{ii} &=& \sum_{j\neq i}(\ba_1)_{ii}(\bb\ba_2\bb^T)_{ij}(\ba_1)_{jj}(\ba_3)_{ji} \:,
\end{eqnarray}
which lead to $(\ba_1\ba_3\bX)_{ii} = (\bX\ba_1\ba_3)_{ii}$ as $\ba_1$, $\ba_2$ and $\ba_3$ are symmetrical matrices. Therefore we find that
\begin{eqnarray}
(\bV_1)_{ii} =&& \left[ \sum_{n=0}^{+\infty}(-1)^n(\ba_1\ba_3)^n + \sum_{n=0}^{+\infty}(1+f_n(\ba_1\ba_3))\bX^n \right]_{ii}(\ba_1)_{ii}\:, \label{eq:V} \\
(\bV_1)_{ii} \approx && \left[ \sum_{n=0}^{3}(-1)^n[(\ba_1\ba_3)^n]_{ii} + \sum_{n=0}^{+\infty}(\bX_{ii})^n \right](\ba_1)_{ii}  \label{eq:V0} \\
= && \left[ \sum_{n=0}^{3}(-1)^n[(\ba_1\ba_3)^n]_{ii} + \frac{\bX_{ii}}{1-\bX_{ii}} \right](\ba_1)_{ii}  \:. \label{eq:V1}
\end{eqnarray}
Here $f_n(\ba_1\ba_3)$ is a power series about $\ba_1\ba_3$, for example, $f_1(\ba_1\ba_3) = \sum_{i=1}^{+\infty}(-1)^i(1+i)(\ba_1\ba_3)^i$.  $f_n(\ba_1\ba_3)\bX^n$ (and the non-diagonal elements of $\bX$) represent the mixing contributions from MC statistical uncertainty and the correlation between signal shape and systematical variation or the correlation between different systematical uncertainties. To derive Eq.~\ref{eq:V}, we have to assume that both MC statistical uncertainty and these correlations are small so that it is valid to represent $\bV$ as a series. Because MC statistical uncertainty and other systematical uncertainty are usually independent, it is difficult to keep terms in a consistent way if do not know their sizes. From Eq.~\ref{eq:V} to Eq.~\ref{eq:V0}, we keep the leading terms considering the correlation between different systematical uncertainties and the dominant terms considering MC statistical uncertainty and omit the mixing contributions, which seems reasonable. 
   
The main results, namely, Eq.~\ref{eq:dmu_Nsys} and Eq.~\ref{eq:dalpha_Nsys}, are derived from the Eq.~\ref{eq:V1}. In practice, we care more about the systematic uncertainty source than the MC statistical uncertainty (if MC statistical uncertainty is dominant, we will usually resort to data-driving methods or increase the MC statistics).  To estimate the precision of $\hatdmu$ using Eq.~\ref{eq:dmu_Nsys}, we can investigate the term, $(\ba_1\ba_3)^4\ba_1$ (In fact, it is not difficult to write down the expression for a general term $(\ba_1\ba_3)^n\ba_1$ by induction).
\begin{eqnarray}
[(\ba_1\ba_3)^4]_{11}(\ba_1)_{11} = && \sum_{i\neq j}\frac{(s\otimes\Delta^i)^2}{s\otimes s(1+\Delta^i\otimes\Delta^i)}\frac{(s\otimes\Delta^j)^2}{s\otimes s(1+\Delta^j\otimes\Delta^j)} \nonumber \\
+ && \sum_{i\neq j, i\neq k, j\neq k} \frac{(s\otimes\Delta^i) (\Delta^i\otimes\Delta^j) (\Delta^j\otimes\Delta^k) (s\otimes\Delta^k)}{s\otimes s(1+\Delta^i\otimes\Delta^i)(1+\Delta^j\otimes\Delta^j)(1+\Delta^k\otimes\Delta^k)} 
\end{eqnarray}
Obviously, they are subleading contributions compared to Eq.~\ref{eq:dmu_Nsys}. They are important only when there are multiple systematical uncertainties which are highly correlated with each other or with the signal shape. For the precision of Eq.~\ref{eq:dalpha_Nsys} to calculate $\hat{\sigma}_{\alpha_i}$, we can look at the omitted sub-leading terms.
\begin{equation}
-2\sum_{j\neq i} \frac{(s\otimes\Delta^i)(s\otimes\Delta^j)(\Delta^i\otimes\Delta^j)}{s\otimes s(1+\Delta^i\otimes\Delta^i)(1+\Delta^j\otimes\Delta^j)} - \sum_{j\neq i, k\neq i, k\neq j} \frac{(\Delta^i\otimes\Delta^j)(\Delta^i\otimes\Delta^k)(\Delta^j\otimes\Delta^k)}{(1+\Delta^i\otimes\Delta^i)(1+\Delta^j\otimes\Delta^j)(1+\Delta^k\otimes\Delta^k)}
\end{equation}
They are minor contributions compared to Eq.~\ref{eq:dalpha_Nsys}.  Using the pseudo experiment described in Sec.~\ref{sec:example}, we confirm that considering these terms will pull the calculated results closer to the results from the fitting tool. But the improvement is limited, and we do not want to include them to make the formulae too cumbersome.

\end{appendix}

\end{document}